\documentclass[fleqn,usenatbib]{mnras}
\usepackage[LGR,T1]{fontenc}
\usepackage[utf8]{inputenc}

\usepackage[british]{babel} 
\usepackage[autostyle, autopunct, english=british]{csquotes}

\def\mytitle{The dependence of protostar formation on the geometry and strength of the initial magnetic field}
\def\myauthor{B. T. Lewis and M. R. Bate}
\def\myrunhead{Effect of field geometry on protostellar outflows}
\def\mykeywords{accretion, accretion discs --- binaries: general --- MHD --- stars: formation --- stars: jets --- stars: winds, outflows.}
\def\mykeywordspdf{accretion, accretion discs, binaries: general, MHD, stars: formation, stars: jets, stars: winds; outflows} 

\hypersetup{pdfauthor={\myauthor}, pdftitle={\mytitle}, pdfkeywords={\mykeywordspdf}, pdfsubject={\myrunhead}}

\def\mylbp{\citet{2015MNRAS.451.4807L}}

\usepackage{xspace}

\usepackage{amsmath}

\usepackage{txfonts}
 \let\la=\lesssim 
\usepackage[cal=rsfso,calscaled=.96,scr=rsfs,scrscaled=.96]{mathalfa}

\usepackage{graphicx}
\makeatletter
\def\ScaleIfNeeded{%
  \ifdim\Gin@nat@width>\linewidth
    \linewidth
  \else
    \Gin@nat@width
  \fi
}
\makeatother

\usepackage{nicefrac}

\usepackage{cleveref}
\Crefname{figure}{Figure}{Figures}
\crefname{figure}{figure}{figures}
\Crefname{equation}{Equation}{Equations}
\crefname{equation}{equation}{equations}

\usepackage{aas_macros}
\usepackage{textcomp}
\usepackage{gensymb}%
\usepackage{empheq}

\usepackage{setspace}

\def\diff{} 

\title[\myrunhead{}]{\mytitle{}}    
\author[\myauthor{}]{Benjamin T. Lewis\thanks{E-mail:
\href{mailto:blewis@astro.ex.ac.uk}{blewis@astro.ex.ac.uk}} and Matthew R. Bate\\
School of Physics and Astronomy, University of Exeter, Stocker Road, Exeter EX4 4QL}

\input{macros}
\renewcommand{\micromu}{\textmu}

\begin{document}

\date{Accepted 2017 January 30. Received 2017 January 27; in original form 2016 January 18.}

\pagerange{\pageref{firstpage}--\pageref{lastpage}} \pubyear{2015}

\maketitle

\label{firstpage}

\begin{abstract}
We report results from twelve simulations of the collapse of a molecular cloud core to form one or more protostars, comprising three field strengths (mass--to--flux ratios, $\mu$, of 5, 10, and 20) and four field geometries (with values of the angle between the field and rotation axes, $\vartheta$, of $0\degree$, $20\degree$, $45\degree$, and $90\degree$), using a smoothed particle magnetohydrodynamics method. We find that the values of both parameters have a strong effect on the resultant protostellar system and outflows. This ranges from the formation of binary systems when $\mu = 20$ to strikingly differing outflow structures for differing values of $\vartheta$, in particular highly suppressed outflows when $\vartheta = 90\degree$. Misaligned magnetic fields can also produce warped pseudo-discs where the outer regions align perpendicular to the magnetic field but the innermost region re--orientates to be perpendicular to the rotation axis. We follow the collapse to sizes comparable to those of first cores and find that none of the outflow speeds exceed $8~\kvelo$.
These results may place constraints on both observed protostellar outflows, and also on which molecular cloud cores may eventually form either single stars and binaries: 
a sufficiently weak magnetic field may allow for disc fragmentation, whilst conversely the greater angular momentum transport of a strong field may inhibit disc fragmentation.
\end{abstract}


\section{Introduction}
\label{sec:intro}
Molecular clouds, and hence the molecular cloud cores which collapse to form protostars, are magnetised \citep{1993ApJ...407..175C,2012ARA&A..50...29C}, as are protostars and the structures surrounding them. In particular, magnetic outflows are proposed as a mechanism to remove angular momentum from protostars --- some mechanism to do so is necessary since simulations without magnetic fields \citep{1978ApJ...224..497N,1994MNRAS.271..999B} and (semi-)analytic calculations \citep[e.g.][]{1984ApJ...286..529T} show that the total angular momentum of cores are too high to form stars without some angular momentum transport mechanism, and \citet{1986ApJ...309..275H} find that the rotation rates of T Tauri stars are substantially lower than the energy required to break up the stellar core. If, for certain field strength or geometries, these outflows can be altered or suppressed, this will constrain how star formation proceeds.

Conversely, some cores must fragment to produce binaries, while still removing sufficient angular momentum to form stellar cores which do not break up. By varying the field strength a regime whereby some magnetic braking, plus a magnetic outflow, removes enough angular momentum to allow stellar core formation while still forming a binary may be obtained. 

 Misalignment between magnetic fields and outflows have been observed on scales of molecular clouds \citep[e.g.][]{2014Natur.514..597S,2013AAS...22142604H,2013ApJ...768..159H} down to stellar length-scales, \citep[e.g.][]{2010MNRAS.409.1347D}. A range of mass-to-flux ratios are observed in molecular clouds, ranging from the observation by \citet{2005IAUS..227...98C} that the mean mass-to-flux ratio, expressed in terms of the critical value, for \textit{massive} star--forming regions is approximately unity, to \citet{2010ApJ...724L.113B} and \citet{2013A&A...556A..16G}. The latter found a much larger range of mass-to-flux ratios in star forming regions, including both sub- and super-critical clouds, a result reinforced by the Bayesian analysis of \citet{2010ApJ...725..466C}. 

From its invention \citep{1977AJ.....82.1013L,1977MNRAS.181..375G} the smoothed particle hydrodynamics (SPH) method has been applied to star formation. In \mylbp{} we used a smoothed particle magnetohydrodynamics (SPMHD) method to model the collapse of a molecular cloud cores to the scale of first hydrostatic cores \citep{1969MNRAS.145..271L}, and the nature of their outflows, with the magnetic field both aligned and misaligned to the rotation axis. We found a significant dependence on the angle between the initial magnetic field and rotation axes (hereafter, $\vartheta$) and the nature of the outflow produced (or, indeed, whether an outflow is produced at all). 
This result was similar to that obtained by \citet{2010MNRAS.409L..39C} (who used a Godunov adaptive mesh refinement code rather than SPH), albeit with some differences. In particular, they found that the mass ejection rate reduced as $\vartheta$ was increased, subsequently increasing the protostellar accretion rate, while we observed that for $20\degree \leq \vartheta \leq 60\degree$ the accretion rate decreased.

\citet{2015MNRAS.451.4807L} adopted an initial magnetic field strength corresponding to a mass-flux ratio of $\mu = 5$. While mass--to--flux ratios close to the critical value are observed, it is instructive to test the effects of weaker fields. Previous work has indicated that the nature of the outflow in an aligned model ($\vartheta = 0\degree$) changes with magnetic field strength. This is to be expected --- purely hydrodynamic simulations do not produce outflows, so the nature and strength of these outflows must be related to the strength of the magnetic field \citep[see, for example,][]{2003MNRAS.339.1223P}. \citet{2014MNRAS.437...77B} found that whilst the velocity of the first-core outflow is essentially unchanged, the width of the outflow (i.e. the opening angle) increases when the initial field is weaker. In addition, stronger and weaker fields will affect the initial collapse of the molecular cloud core (and may prevent it collapsing at all if a sufficient magnetic pressure can be realised). 

Therefore, in this work we perform a similar series of calculations to \mylbp{}, but with weaker fields --- corresponding to mass-flux ratios of $\mu = 10$ and $20$. We present SPMHD simulations of three mass--to--flux ratios and four field geometries. This allows us to probe a range of field strengths. Additionally we can observe how the changing field geometries changes the nature and extent of the outflows observed. We do not include non--ideal MHD effects \citep[see][]{2007Ap&SS.311...35W,2014MNRAS.444.1104W,2016MNRAS.457.1037W,2015MNRAS.452..278T,2015ApJ...810L..26T} and therefore can not probe the regime where $\mu$ is less than or equal to unity.

\Cref{sec:method} details our numerical method, which is essentially that used in \citet{2014MNRAS.437...77B} and \cref{sec:initial} details the initial conditions. Then in \cref{sec:isothermal,sec:discs:5,sec:discs:10,sec:discs:20} we present the results of all twelve models and a discussion of the conclusions obtained from them. We then consider the process of accreting material onto the sink particles in \cref{sec:accretion}, before finally comparing our results to some observations in \cref{sec:obs}.

\section{Method}
\label{sec:method}

We solve the equations of ideal magnetohydrodynamics with gravity, viz.
\begin{equation}
\frac{\rmn{d}\rho}{\rmn{d}t} = -\rho{}\nabla{}^{i}v^{i} ~\text{,}
\end{equation}
\begin{equation}
\frac{\rmn{d}v^{i}}{\rmn{d}t} = \frac{1}{\rho} \nabla^{j} S^{ij} - \nabla^{i}\phi{} ~\text{,}
\label{eqn:mom_analyitc}
\end{equation}
\begin{equation}
\frac{\rmn{d}B^{i}}{\rmn{d}t}  = \left(B^{j}\nabla^{j} \right)v^{i} - B^{i} \left(\nabla^{j} v^{j} \right) ~\text{,}
\end{equation}
\begin{equation}
\nabla^{2}\phi{} = 4\pi{}G\rho{} ~\text{,}
\end{equation}
with the MHD stress tensor
\begin{equation}
S^{ij} = - P \delta^{ij} +\frac{1}{\mu_0} \left( B^i B^j - \frac{1}{2}\delta^{ij}B^2 \right) \text{~.}
\end{equation}
In these equations and the rest of the paper, we represent the total derivative by
\begin{equation} 
\frac{\rmn{d}~}{\rmn{d}t} = \frac{\partial{}~}{\partial{}t} + v^{i}\nabla^{i} ~\text{,}
\end{equation}
the symbols $P$, $\rho$, $v^{i}$, $B^{i}$, and $\phi$ represent the pressure, density, velocity, magnetic and gravitational fields respectively, $\delta^{ij}$ is the Kronecker delta, $\mu_0$ is the permeability of free space, and repeated indices imply summation.

We use \textsc{sphng}, a three--dimensional hybrid \textsc{openmp} and \textsc{mpi} smoothed particle radiation magnetohydrodynamics code that originated from a code written by \citet{X}, with subsequent extensive modifications by \citet{1995MNRAS.277..362B} and others.
We discretise these equations using the SPMHD method derived originally in 
\citet{2004MNRAS.348..123P} \citep[see also the review by ][]{2012JCoPh.231..759P} and subsequently modified as detailed below. We use the 
source term correction to the tensile instability proposed by \citet{2001ApJ...561...82B}, with the parameter $\chi = 1$ \citep[see][]{LBT2016}. Each particle has an individual smoothing length, $h_a$, given by
\begin{equation}
h_a = \eta \left( \frac{m_a}{\rho_a} \right)^{\frac{1}{\nu}} \text{~,}
\end{equation}
where the symbol $a$ denotes individual SPH particles, $m_a$ represents the mass of an SPH particle -- which is identical for all SPH particles in our model, $\nu = 3$ is the number of spatial dimension and $\eta = 1.2$ for the cubic spline kernel \citep{1985A&A...149..135M}.

To capture shocks and prevent inter--particle penetration we use artificial viscosity and resistivity terms following \citet{1997JCoPh.136..298M}, \citet{2004MNRAS.348..123P} and \citet{2005MNRAS.364..384P}. Both artificial dissipation terms are controlled by switches which are spatially and temporally variable, for artificial viscosity we use the \citet{1997JCoPh.136...41M} switch with $\alpha_{\text{AV}} \in [0.1, 1.0]$ while for artificial resistivity the \citet{2013MNRAS.436.2810T} switch with $\alpha_{\text{B}} \in [0.0, 1.0]$.

The gravitational force is calculated using the method in \citet{2007MNRAS.374.1347P} with the potential softened using the SPH smoothing kernel, where we use the same binary tree to both find neighbours and to compute the gravitational force.

To maintain the solenoidal constraint on the magnetic field (since $\nabla^i B^i = 0$ only enters the equations of MHD as an initial condition and $\nicefrac{\rmn{d}~}{\rmn{d}t} \nabla^i B^i$ is not guaranteed to be zero) we use the hyperbolic divergence cleaning derived by \citet{2012JCoPh.231.7214T} \citep[based on the earlier Eulerian method of][]{2002JCoPh.175..645D}. We set the damping time-scale,
\begin{equation}
\tau = \frac{h}{\sigma c_\rmn{c}} \text{~,}
\end{equation}
such that the cleaning wave is critically damped by choosing $\sigma = 0.8$, and set the cleaning wave speed to the maximum magnetosonic wave speed, i.e.
\begin{align}
\begin{split}
c_\rmn{c}^2 &= c_\rmn{s}^2 + v_\rmn{a}^2 = c_\rmn{s}^2 + \frac{B^2}{\mu_0 \rho} \text{~.}
\end{split}
\end{align}
These calculations were performed using the original cleaning method, as opposed to the modified scheme using $\nicefrac{\psi}{c_\rmn{c}}$ recently proposed by \citet{2016JCoPh.322..326T}.

Each SPH particle is assigned an individual timestep based on the local Courant condition. The equations are discretised in time using a second--order Runge--Kutta--Fehlberg integrator \citep{RK4NASATR}. Depending on the exact parameters employed, each calculation ran for between ca. 100 and 400 hours of wall time on two 6-core hyper-threaded CPUs (i.e. 24 execution threads, and therefore between 2,400 and 9,600 hours of computation-core time).

The integrator bug that affected the calculations of \citet{2015MNRAS.451.4807L}, reported in the Erratum and Addendum to that paper \citep{2017MNRAS.464.2499L}, was fixed \textit{before} the calculations for this paper were performed. Consequently, we also do not use the `average $h$' terms that were used in that previous work.

We add sink particles to the simulation once a critical density of $\rho_\rmn{crit} = 10^{-10}~\udens$ is exceeded and the usual tests (see \citet{1995MNRAS.277..362B} for details) are passed. By doing so we are able to avoid resolving the formation of the stellar core itself and therefore can evolve the simulations further. The accretion radius of the sink particle is set to 1 \au, any SPH particle that comes within this distance of the sink and passes the \citet{1995MNRAS.277..362B} tests will be removed from the simulation and its mass will be added to the sink particle. Other than through gravity, sink particles do not interact in the simulation, in particular they do not possess a hydrodynamic pressure nor a magnetic field.

\section{Initial Conditions}
\label{sec:initial}

The initial conditions are essentially the same as in previous work, which ranges from the early work of \citet{HOSKING2002}, and the Euler potential SPMHD method in \citet{2007MNRAS.377...77P}, to the self-consistent direct induction SPMHD method in \citet{2012MNRAS.423L..45P,2014MNRAS.437...77B,2015MNRAS.451.4807L}. Calculations with different set ups have also been performed, for example, \citet{2006ApJ...641..949B} who used Bonner-Ebert \citep{1956MNRAS.116..351B,1955ZA.....37..217E} spheres instead of the uniform density sphere used here. An $M_\rmn{sphere} = 1~\solarm$ sphere of cold gas was placed in a warm external medium. The sphere and medium are in pressure equilibrium, however, the sphere is 30 times the density of the medium, therefore the initial temperatures are $\approx 10~\kelvin$ (corresponding to an initial isothermal sound speed of $c_\rmn{s} = 2.2\times10^4~\velo$) and $\approx 300~\kelvin$ respectively. The sphere had an initial radius $r_0 = 4\times 10^{16}~\cm$ and therefore an initial density of $\rho_0 = 7.4 \times 10^{-18}~\udens$. The sphere was then placed in a cubic periodic box with side lengths $16 \times{} 10^{16}~\cm$ to prevent numerical artefacts in the magnetic field at the edge of the warm medium.

An effective minimum resolution is provided by the \citet{1997MNRAS.288.1060B} criterion, namely that the Jeans length must be resolved throughout the collapse of the core, which would require ca. 30,000 particles; however, \citet{2011ApJ...731...62F} find that magnetised turbulence may require a resolution well in excess of this. \citet{2014MNRAS.437...77B} find no significant resolution dependence between 1 and 10 million particles; consequently we use 1.5 million SPH particles in the core as a compromise between computational expense and numerical accuracy. The sphere is then surrounded by ca. 700,000 particles to represent the warm medium. Aside from providing a \enquote{container} for the sphere and providing a self-consistent magnetic boundary, the warm medium is dynamically uninteresting --- if the sphere was instead placed in vacuo similar results would be expected, albeit with a slower initial collapse \citep[which was inadvertently shown in][who accidentally performed their calculations with a pressure-free external medium]{2007MNRAS.377...77P}. 

We used a barotropic equation of state similar to \citet{2008ApJ...676.1088M}, without the final $\gamma = \nicefrac{5}{3}$ step,
\begin{empheq}[left={P = c^{2}_\rmn{s} \empheqlbrace}]{equation}
\begin{aligned}
& \rho{}          &&\rho{} \leq \rho_{\rmn{c,1}} \\
& \rho_{\rmn{c,1}} \left( \frac{\rho{}}{\rho_{\rmn{c,1}}} \right)^{\frac{7}{5}} & \hspace{-5pt}\rho_{\rmn{c,1}} <~ & \rho{} \leq \rho_{\rmn{c,2}} \\
& \rho_{\rmn{c,1}} \left( \frac{\rho_{\rmn{c,2}}}{\rho_{\rmn{c,1}}} \right)^{\frac{7}{5}} \left( \frac{\rho{}}{\rho_{\rmn{c,2}}} \right)^{\frac{11}{10}} && \rho{} > \rho_{\rmn{c,2}}
\end{aligned}
\end{empheq}
where the critical densities are given by $\rho_\rmn{c,1} = 10^{-14}~\udens$ and $\rho_\rmn{c,2} = 10^{-10}~\udens$ (the physical consequences of these choices are discussed in \cref{sec:discs}).

The magnetic field is characterised by the dimensionless mass-to-flux ratio, $\mu$, and the angle between the rotation and field axes, $\vartheta$. The mass-to-flux ratio, in terms of the critical value, is defined as
\begin{equation}
\mu = \frac{\varpi_\rmn{sphere}}{\varpi_\rmn{critical}}
\end{equation}
where the ratio between the magnetic and gravitational forces in a magnetised sphere is given by
\begin{equation}
\varpi_\rmn{sphere} = \frac{M_\rmn{sphere}}{\rmn{\pi}r^2_0 B_0}
\end{equation}
and the critical ratio, i.e. that at which the self-gravitational and magnetic pressure forces will be in equilibrium, is given by
\begin{equation}
\varpi_\rmn{critical} = \frac{2c_1}{3}\sqrt{\frac{5}{\rmn{\pi}G\mu_0}}.
\end{equation}
where the parameter $c_1 = 0.53$ \citep{1976ApJ...210..326M}. The geometry of the initial field, $B_0$ is then defined according to
\begin{equation}
B^{x}_0 = B_0 \sin\vartheta \text{~,}
\end{equation}
\begin{equation}
B^z_0 = B_0 \cos\vartheta \text{~,}
\end{equation}
so that when $\vartheta = 0\degree$ the field and rotation axes are parallel (i.e. aligned) and, conversely, when $\vartheta = 90\degree$ they are perpendicular (i.e. totally misaligned). The rotation axis is aligned with the $z-$axis, so that the magnetic field axis then varies. Owing to the symmetry of the initial field -- and the fact the field in the warm medium does not materially evolve over the course of the calculation -- the field is also periodic across the boundaries, preventing numerical artifacts at the box edge.

We use mass-to-flux ratios of $\mu = [5,~10,~20]$, which give field strengths of $B_0 = [163,~81,~41]$ microgauss and mean ratios of hydrodynamic to magnetic pressure, $\beta = [3.45,~13.8,~55.3]$ respectively (i.e. in all cases the simulation begins with the hydrodynamic pressure stronger than the magnetic pressure). For each field strength, we then perform the calculations 4 times, with values of $\vartheta = [0\degree,~20\degree,~45\degree,~90\degree]$ to give a total of 12 models.

Finally, the sphere is placed in solid body rotation around the $z$-axis with $\Omega = 1.77 \times 10^{-13}~\rads$, which gives a ratio of rotational to gravitational energy of $\beta_\rmn{rot} = 0.005$, within the range of values observed by \citet{1993ApJ...406..528G}.

\section{Results and Discussion}
\label{sec:discs}

\subsection{Isothermal Collapse}
\label{sec:isothermal}

\begin{figure}
\centering{}
\includegraphics{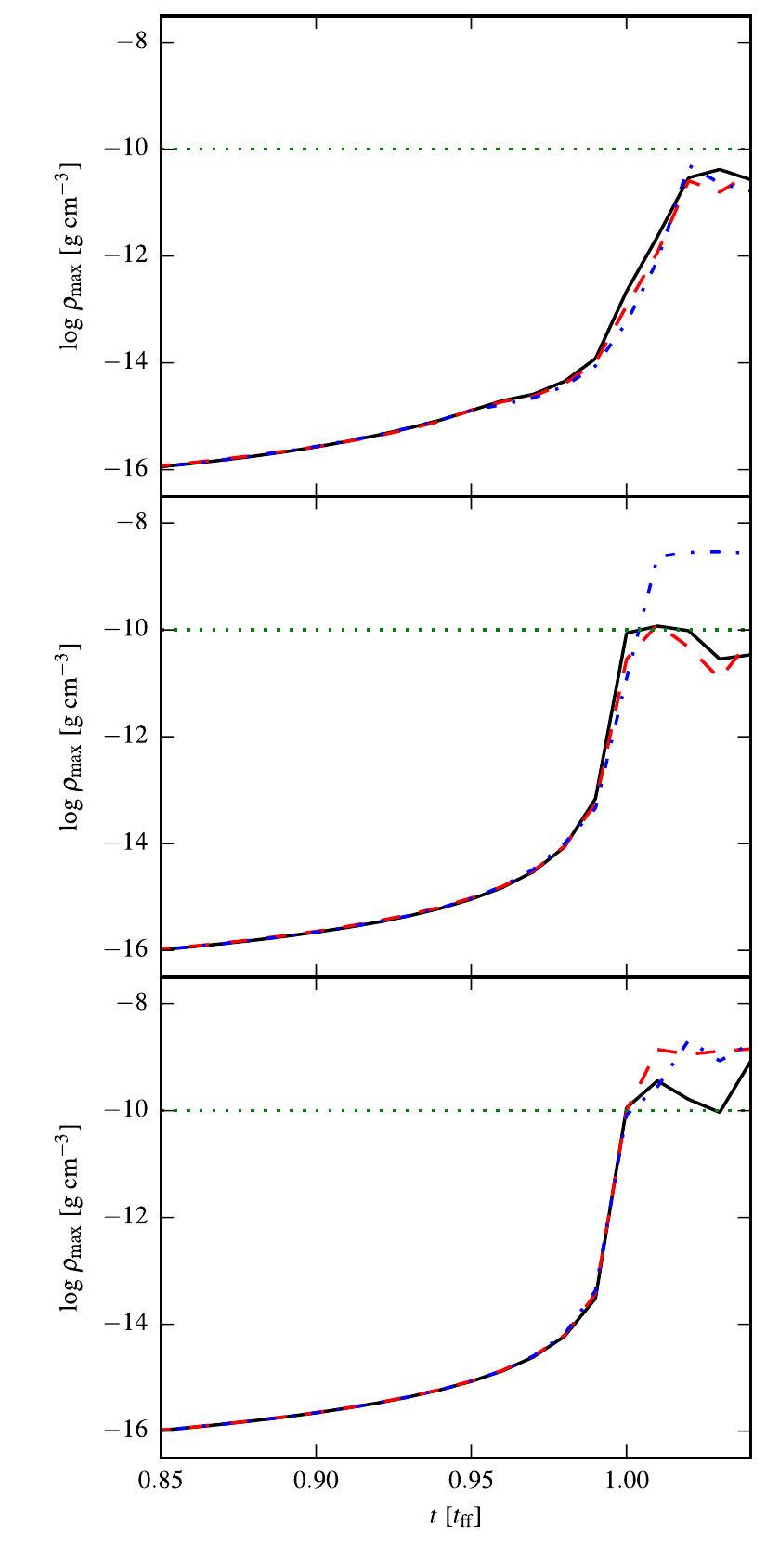}

\caption{Maximum density as a function of time (where one free--fall time, $\tff = 24\thinspace430~\yr$) for each of $\mu = 5$ (solid black line), $10$ (dashed red line), and $20$ (dot--dashed blue line) models with $\vartheta = 0\degree$ (upper panel), $45\degree$ (middle panel), and $90\degree$ (lower panel) respectively. In all cases, the $\mu = 10$ and $20$ calculations reach $\rho_\rmn{max} > 10^{-11}\udens$ $0.01\tff$ earlier than the $\mu = 5$ calculations.
However, until $t \approx 0.97t_\rmn{ff}$ the maximum densities are similar. The dotted line at $\rho_\rmn{crit} = 10^{-10}~\udens$ indicates the critical density above which sink particles may be inserted if the appropriate formation tests are passed, and is therefore an approximate maximum density for the calculation. In the $\mu = 5$ calculations, the sink insertion quickly removes the entire region where $\rho > \rho_\rmn{crit}$ and consequently the maximum density appears to not exceed the sink formation threshold; for both $\mu = 10$ and $20$ a region where $\rho > \rho_\rmn{crit}$ persists and is not removed by the insertion of further sink particles since one or more of the creation tests in \citet{1995MNRAS.277..362B} has failed. This high density gas exists in a disc around the sink particle.
\label{fig:rhotime}}
\end{figure}

The sphere of cold gas described in \cref{sec:initial} is allowed to collapse. For this discussion, we divide the collapse of the cloud core into two phases: firstly an initial isothermal collapse, where the maximum densities are below $\rho = \rho_\rmn{c,1} = 10^{-14}~\udens$ and then a second phase where a disc structure and pre-stellar core is formed. We first describe the initial isothermal phase which, aside from slight variations in timings, is comparable for all three field strengths. We then, in \cref{sec:discs:5,sec:discs:10,sec:discs:20}, consider the three field strengths in more detail, focusing in \cref{sec:discs:5,sec:discs:10} on the formation of bipolar outflows and how these vary with the field geometry, and in \cref{sec:discs:20} on how the weakest field strength can form binary and multiple protostellar systems.

\begin{figure}

\centering{}
\includegraphics[width=0.5\textwidth]{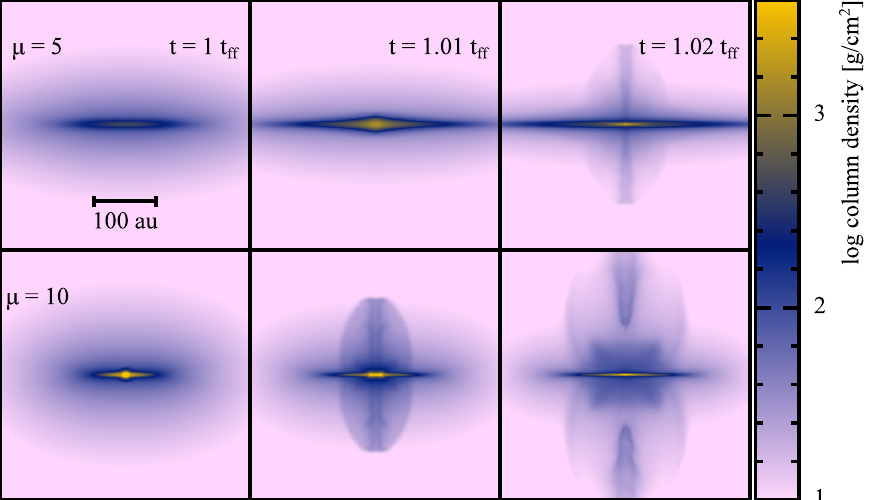}
\caption{Column density projections in the $z$--$x$ plane for both the $\vartheta = 0\degree$ $\mu = 5$ and $10$ models as the sink particle is inserted and jets are formed. The $\mu = 10$ model collapses to a dense disc, and hence forms an outflow, at a dynamical time $\approx 0.01\tff$ earlier than $\mu = 5$.
 \label{fig:sinklaunch}}
\end{figure}

The first phase of the collapse extends from $t = 0$ to $0.98\tff$, corresponding to the region in \cref{fig:rhotime} where the maximum density increases approximately linearly with time (where the free--fall time, $\tff = 24430~\yr$). 
The main effect from the initial conditions observed here is that the lower mass--to--flux ratios provide higher magnetic pressures --- $P_\rmn{mag} \propto B^2$ so the doubling of the field strength between $\mu = 10$ and $5$ provides a corresponding quadrupling of the pressure support. This results in the duration first phase of the collapse being slightly extended for the stronger fields. For example, the $\mu = 10$ calculations reach a density of $10^{-10}~\udens$ $0.01~\tff$ quicker than the stronger field calculations, delaying the subsequent phase of evolution accordingly. \Cref{fig:sinklaunch} shows the formation of the bipolar outflows, which are launched when a pseudo-disc forms and the density exceeds $\approx 10^{-10}~\udens$ (and consequently the $0.01~\tff$ delay when $\mu = 5$) and which we discuss in \cref{sec:discs:5,sec:discs:10}. 

\begin{figure}

\centering{}
\includegraphics[width=0.5\textwidth]{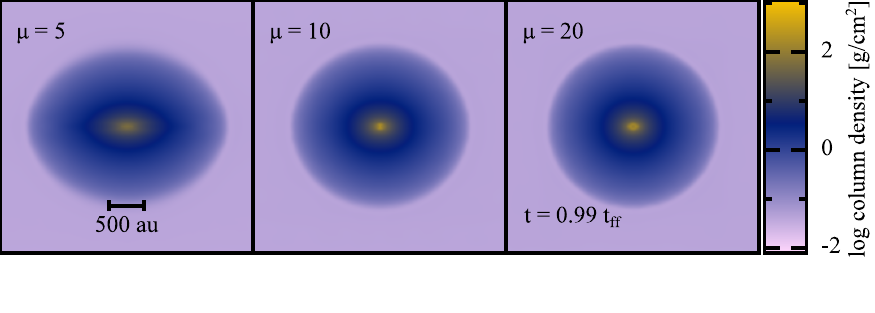}
\caption{Column density projections in the $z$--$x$ plane for $\mu = 5$, 10, and 20 with $\vartheta = 0\degree$ at $t = 0.99\tff$ showing the change in ellipsoidality of the collapsing core. The case with the stronger field, with $\mu = 5$, is much more oblate than those with either $\mu = $10 or 20. We obtain approximate eccentricities of 0.66, 0.34, and 0.24 for each calculation, respectively.
 \label{fig:ellipses}}
\end{figure}

The field geometry, i.e. $\vartheta$, does not affect the overall rate of the collapse since the magnetic pressure terms, which act to support the cloud against gravity, are isotropic and therefore independent of $\vartheta$. However, the tension component \textit{is} anisotropic and, in general, acts to stop the fluid moving perpendicular to the magnetic field lines. Consequently this causes the collapsing core to become ellipsoidal.
Thus, the $\vartheta = 0\degree$ calculations produce an oblate spheroid while conversely the $\vartheta = 90\degree$ calculations produce prolate spheroids and $20\degree$ and $45\degree$ produces a tri-axial ellipsoid. \Cref{fig:ellipses} shows how the oblateness of the spheroids changes with field strength, with the axis ratios of $0.66$, $0.34$, and $0.24$ for $\mu = 5$, $10$, and $20$ respectively.
Equivalent values can be obtained for other geometries (although these are tri--axial ellipsoids or prolate spheroids).

\Cref{fig:rhotime} shows how the maximum fluid density evolves as a function of time for all twelve calculations. At $0.98-0.99\tff$ (depending on the calculation) the density begins to rapidly increase. This marks the change from the isothermal and weakly oblate collapsing core to the second phase. The magnetic forces described above which acted to make the core oblate now cause the formation of a flattened disc-like structure. 
We use the term pseudo--disc to describe this object, which is not \diff{a `classical'} accretion disc \diff{in Keplerian rotation} around a protostellar core.
\diff{N}evertheless \diff{it} is rotating\diff{ albeit with a sub--Keplerian rotation profile,} is supported by the gas pressure perpendicular to the pseudo-disc plane against further gravitational collapse parallel to the magnetic field axis\diff{, and is accreting onto the protostar (i.e. the radial velocity of the disc relative to the protostar is negative). We discuss this further in }\cref{sec:discs:5,sec:discs:10}\diff{.}

\subsection{$\mu = 5$}
\label{sec:discs:5}

\begin{figure*}
\centering{}
\includegraphics{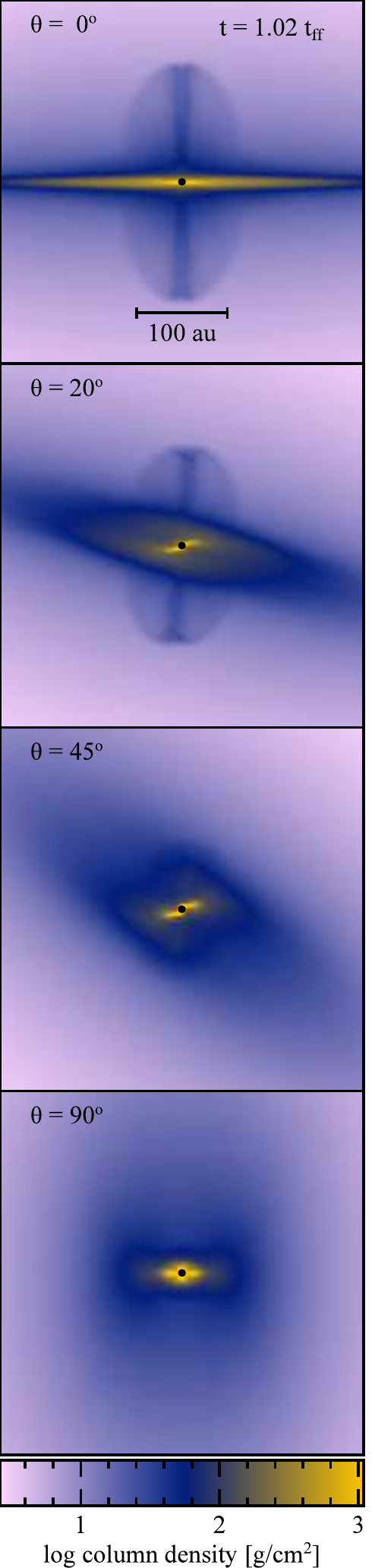}\includegraphics{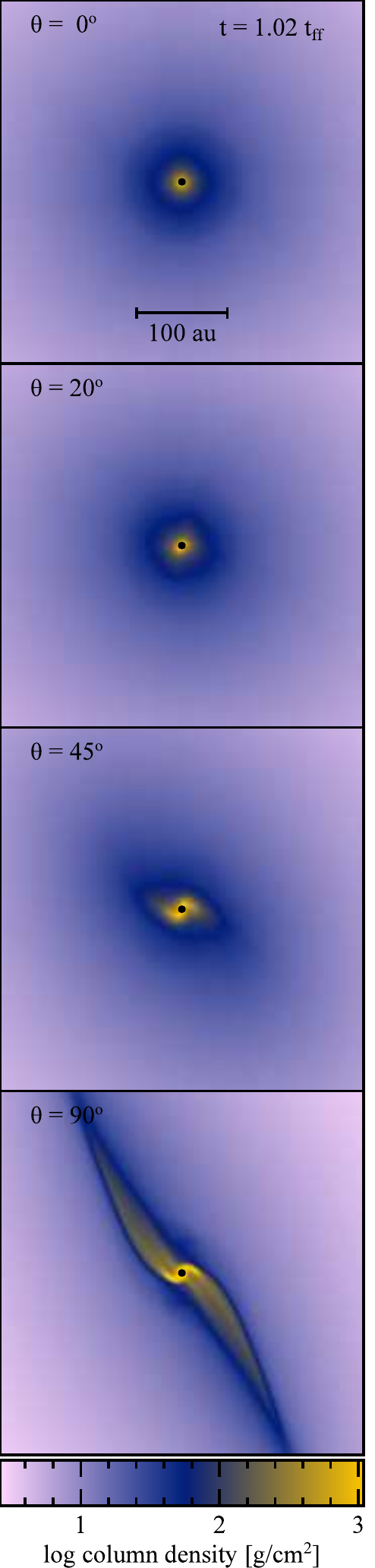}\includegraphics{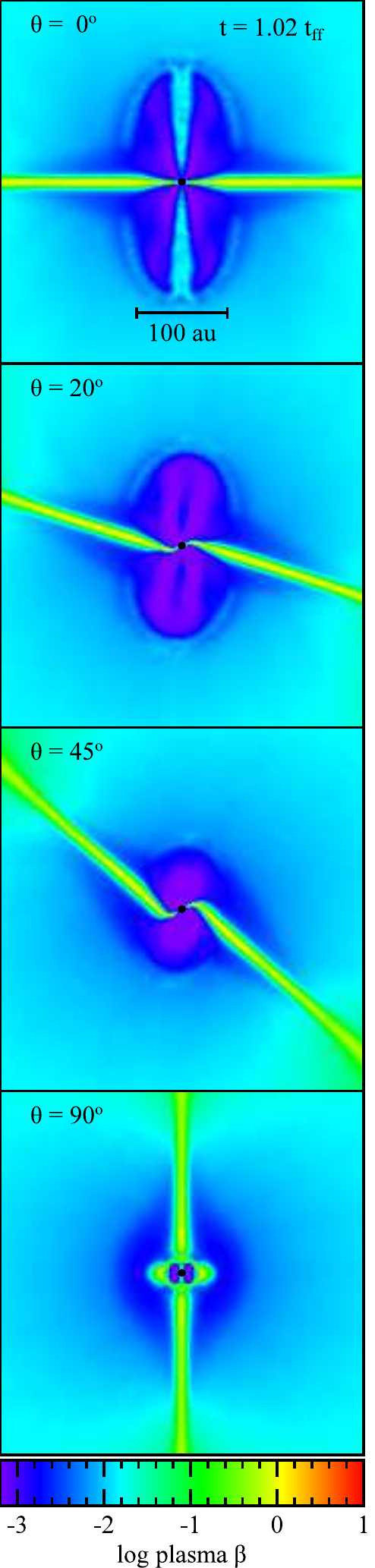}

\caption{\label{fig:mu5all} Column density projections in the direction of the $y$--axis (left
 column) and $z$--axis (centre column), and cross--sections of the plasma $\beta$ in the $z$--$x$ plane
  for the four $\mu = 5$ calculations at $t = 1.02\tff$. The initial inclination between the 
  rotation and magnetic field axes decreases down the page, with the first row being $\vartheta = 0\degree$, 
  then $20\degree$, $45\degree$, and the bottom row having $\vartheta = 90\degree$. 
  \diff{Sink particles are represented by a black dot, approximately four times larger than the accretion radius.}
  The alignment of the large scale pseudo-disc perpendicular to the magnetic field axis can be
  clearly seen, as can the reduction in the outflow velocity and eventual suppression as 
  $\vartheta \rightarrow 90\degree$, as can the co-location of an outflow, i.e. material 
  moving away from the pseudo-disc, and the region where $\beta < 1$ can be seen.
  }
\end{figure*}

\begin{figure*}
\centering{}
\includegraphics[width=\textwidth]{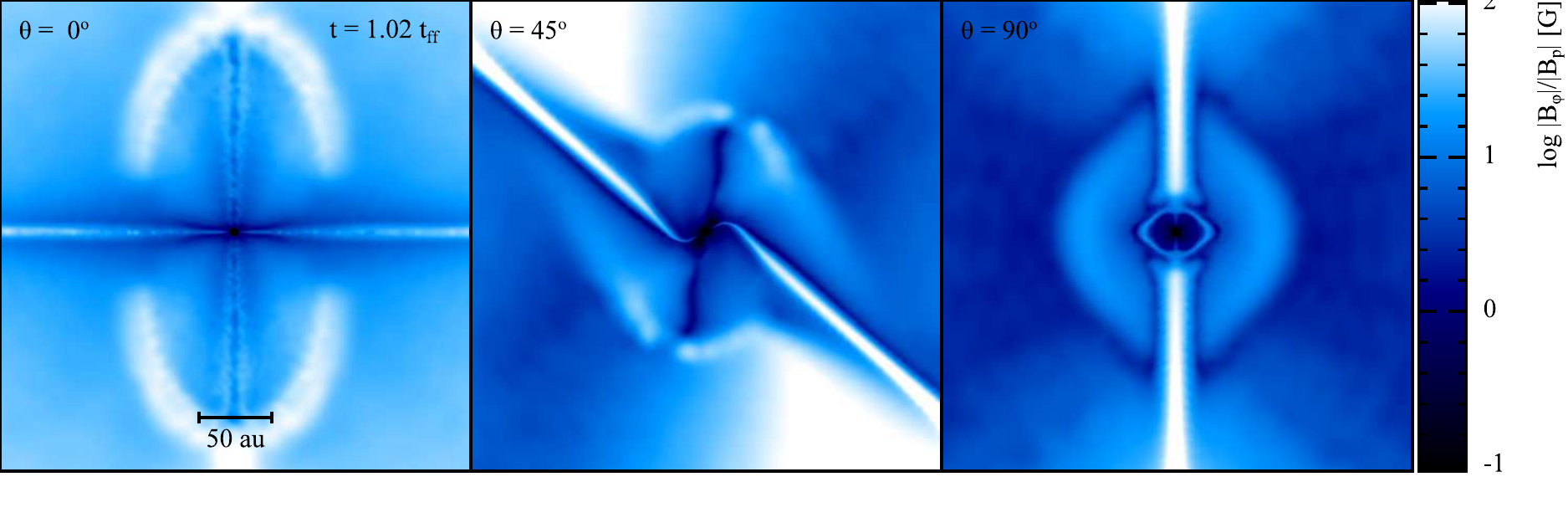}
\caption{\label{fig:TorPol_0} Cross--sections of the ratio of the toroidal field component to the poloidal magnetic field component (\cf \cref{eqn:radialB,eqn:polB,eqn:torB}) for $\vartheta = 0\degree$, $45\degree$, $90\degree$, $\mu = 5$ calculations at $t = 1.02\tff$. Values $< 1$, i.e. black and dark blue, indicate regions where the poloidal field is dominant; conversely values $> 1$ which are indicated with lighter blues and whites are regions where the toroidal field dominates. A clear poloidal region around the sink particle \diff{(approximately a radius of $12~\au$)} in the $\vartheta = 90\degree$ calculation can be seen, as can the strongly toroidal region centered on the rotation axis. This can be contrasted with the other two calculations where the toroidal field (which acts to collimate an outflow) stretches into the pseudo-disc and the region around the sink particle, producing the characteristic bipolar jet.}
\end{figure*}

The strongest initial field strength used in this paper is $\mu = 5$. After the initial isothermal collapse phase discussed above, all four field geometries form small pseudo-discs and bipolar outflows, which we show at $t = 1.02\tff$ in \cref{fig:mu5all}.

Once a significant pseudo-disc forms, magnetic and rotational forces become comparable in magnitude to gravitational forces -- driving the formation of effects like the bipolar outflows. Previous simulations of magnetised collapse and outflows used strong fields, aligned with the rotation axis \citep[\eg][]{2002ApJ...575..306T,2004ApJ...616..266M,2008ApJ...676.1088M}; 
here we also vary $\vartheta$ similar to the approach in \citet{2010MNRAS.409L..39C} and \citet{2015MNRAS.451.4807L}. 

Inclination angles $\vartheta \leq 45\degree$ produce a bipolar outflow. It is convenient to divide these outflows into two regions: an overall bulk outflow and an inner collimated jet. We decompose the magnetic field into cylindrical co-ordinates \citep[for example, as in][]{1955ApJ...122..293P}, such that the (magnitude of the) radial field is given by
\begin{equation}
\label{eqn:radialB}
B_r = \sqrt{B^2_x +B^2_y}
\end{equation}
 the magnitude of the poloidal field as
\begin{equation}
\label{eqn:polB}
B_\rmn{p} = \sqrt{B_r^2 + B^2_z} \comma
\end{equation}
with the azimuthal (i.e. toroidal) field given by
\begin{equation}
\label{eqn:torB}
B_\phi = B_\rmn{tor} = \arctan\thinspace\frac{B_y}{B_z} \fstop
\end{equation}

\Cref{fig:TorPol_0} shows the ratio of the poloidal to toroidal field in a cross--section slice for t the $\vartheta = 0\degree$, $45\degree$, and $90\degree$ calculations.
A thin (\ca $5~\au$ wide) region dominated by the poloidal field may be observed, centred on the rotation axis and hence the jet. This poloidal field acts to move material away from the pseudo-disc which is dominated by the toroidal field. The disc and its toroidal field act to `wind-up' the magnetic field and then material and angular momentum are ejected in the outflow; the central region of which is collimated by the toroidal component. A comparable effect, albeit with a somewhat lesser toroidal field contribution, is seen for $\vartheta=45\degree$ (central panel of \cref{fig:TorPol_0}).

In sharp contrast, the $\vartheta = 90\degree$ calculation has no jet --- but does have a small (with a maximum height above the disc of $10~\au$) bulk outflow. This is produced by the small region around the pseudo-disc which is dominated by the poloidal magnetic field (shown in the right--hand panel of \cref{fig:TorPol_0}). We observe that above this, again centered on the rotation axis, a thin highly toroidal region exists centered on the rotation axis. However, this is not connected to the pseudo-disc and consequently can not act to collimate a jet --- in essence, because the pseudo-disc is rotating in the same plane as the magnetic field lines it is unable to perform the same \enquote{winding} effect seen when $\vartheta = 0\degree$. The presence of a collimated bipolar outflow therefore clearly implies a combination of sufficent magnetic pressure (exemplified by the poloidal field magnitude) to lift material out of the pseudo--disc \textit{combined with} a toroidal field which can collimate and drive the jet \citep[\cf][]{1996MNRAS.279..389L,2001ASPC..249..212L}.

Earlier, we observed that changing the values of $\vartheta$ changed the collapsing cloud core from oblate to tri--axial. This effect continues into the later phase of the collapse where the consequent pseudo--disc aligns perpendicular to the magnetic field axis (and hence parallel to the major axis of the antecedent ellipsoid) \citep{1993ApJ...417..243G}. However, we find that the innermost region of the disc, with a radius of $< 20~\au$, attempts to re--align with the rotation axis, caused by the angular momentum at this scale being sufficient to overcome the magnetic tension force which is aligning the larger scale disc. This produces warped pseudo--discs for $\vartheta > 0\degree$. This effect is most marked when $\vartheta=90\degree$ where the inner disc has a radius of $25~\au$, but can also be seen for $20\degree$ and $45\degree$ albeit with a radii of $5$--$10~\au$, and with a clearer connection to the rest of the pseudo--disc. 

This inner region drives the central collimated region of the outflow. Although the opening angle of the bulk outflow is very large --- the diameter of the whole outflow is on the order of $100~\au$ --- the collimated inner region is much smaller and is correlated with the inner disc. As a result the outflow initially aligns much closer to the rotation axis than the magnetic field axis, for example when $\vartheta = 20\degree$, the inclination of the jet to the rotation axis is less than $10\degree$. The outflow (and in particular the jet) is removing angular momentum from the pseudo--disc so a close alignment between this and the direction of the angular momentum vector is to be expected.

\subsection{$\mu = 10$}
\label{sec:discs:10}

\begin{figure*}
\centering{}
\includegraphics{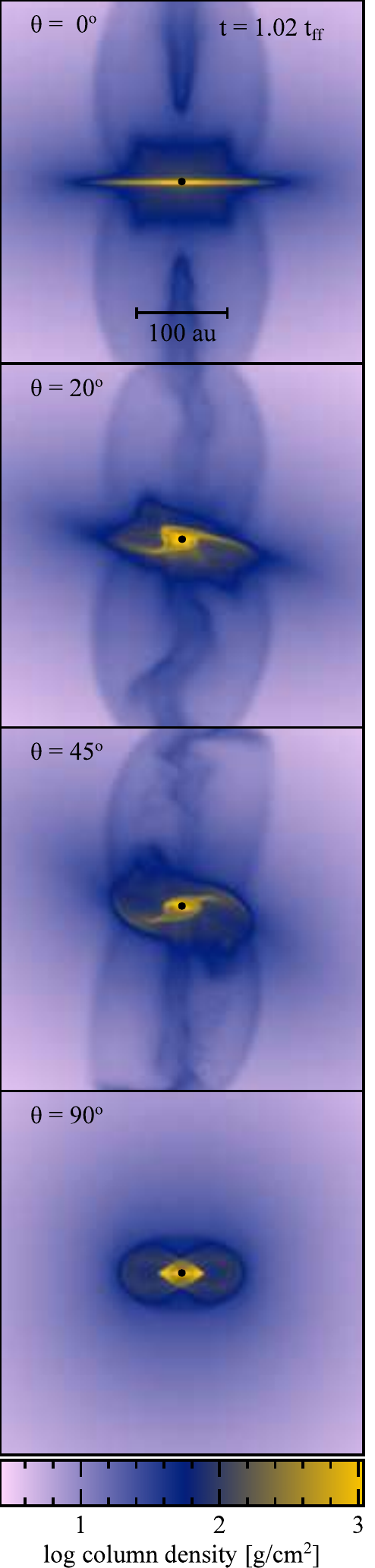}\includegraphics{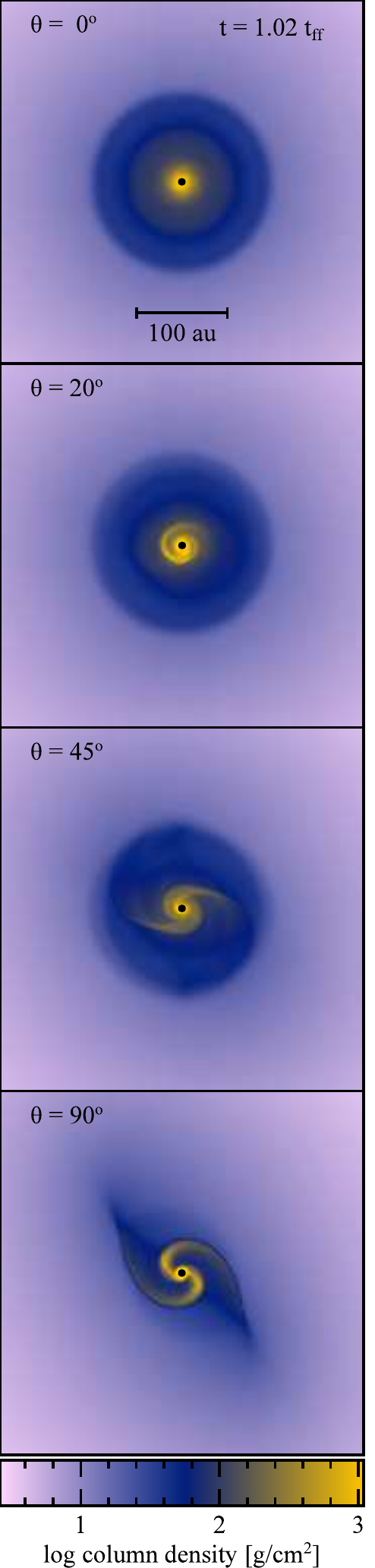}\includegraphics{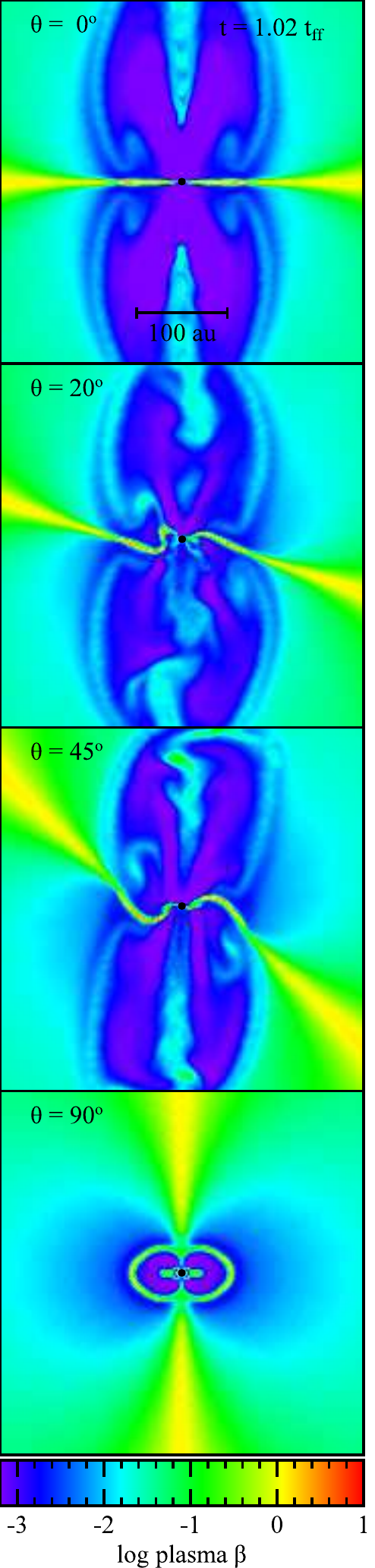}

\caption{\label{fig:mu10all} Column density projections in the direction of the $y$--axis (left
 column) and $z$--axis (centre column), and cross--sections of the plasma $\beta$ in the $z$--$x$ plane
  for the four $\mu = 10$ calculations at $t = 1.02\tff$. 
  \diff{Sink particles are represented by a black dot, approximately four times larger than the accretion radius.} 
  The initial inclination between the 
  rotation and magnetic field axes decreases down the page, with the first row being $\vartheta = 0\degree$, then $20\degree$, $45\degree$, and the bottom row having $\vartheta = 90\degree$. 
  A similar distribution of structures can be seen to those in \cref{fig:mu5all}; however, the pseudo-discs with this weaker initial magnetic field are in general larger due to the reduced angular momentum transport from magnetic braking. The consequently faster rotation rate of the disc produces faster outflow jets, for example the \enquote{disconnection} region when $\vartheta = 0\degree$ has an outflow speed of $|v_z| = 8~\kvelo$.
  }
\end{figure*}

\begin{figure*}
\centering{}
\includegraphics[width=\textwidth]{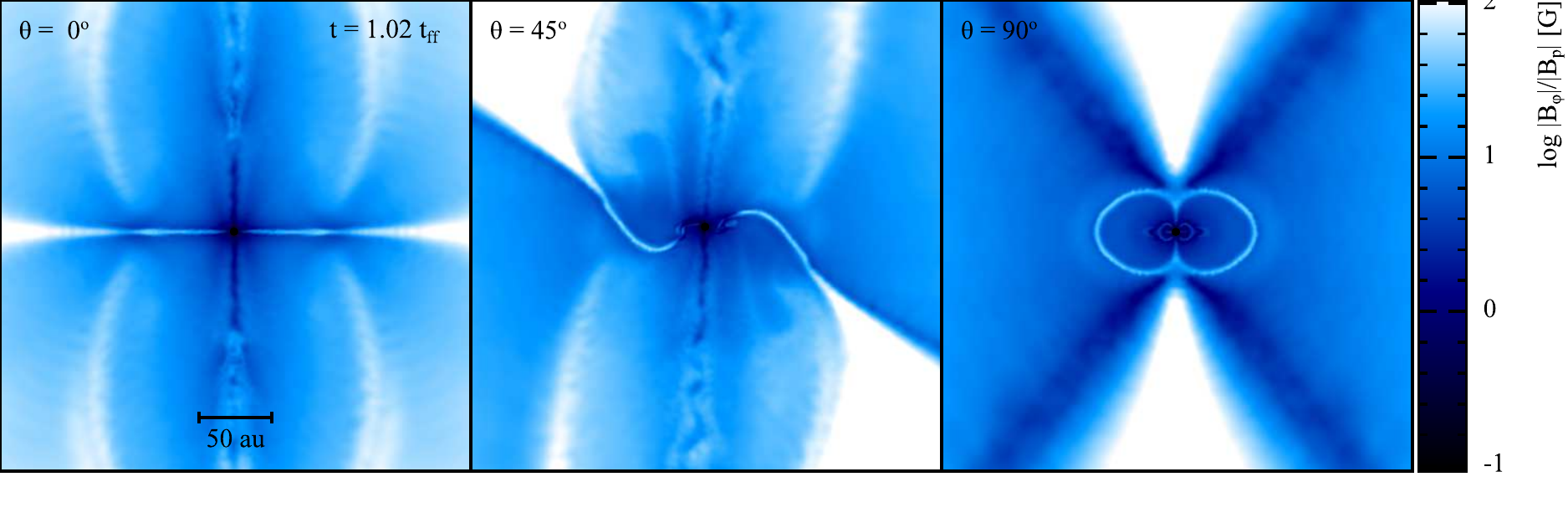}
\caption{\label{fig:TorPol_1} Cross--sections of the ratio of the toroidal field component to the poloidal magnetic field component (cf. \cref{eqn:radialB,eqn:polB,eqn:torB}) for $\vartheta = 0\degree$, $45\degree$, $90\degree$, $\mu = 10$ calculations at $t = 1.02\tff$, similar to \cref{fig:TorPol_0}. Values $<1$ indicate the poloidal component and conversely $>1$ indicates the toroidal component dominates. The \enquote{disconnection} effect in the aligned calculation for $\mu = 10$ (see the top-right panel of \cref{fig:mu10all}) can be attributed to the much larger poloidal region, compared to that seen in \cref{fig:TorPol_0}; $\vartheta = 45\degree$ and $90\degree$ also exhibit larger poloidal regions but these do not cause any marked effect on the jet structure (or in the case of $90\degree$, the lack of it.)}
\end{figure*}

\begin{figure}
\centering{}
\includegraphics{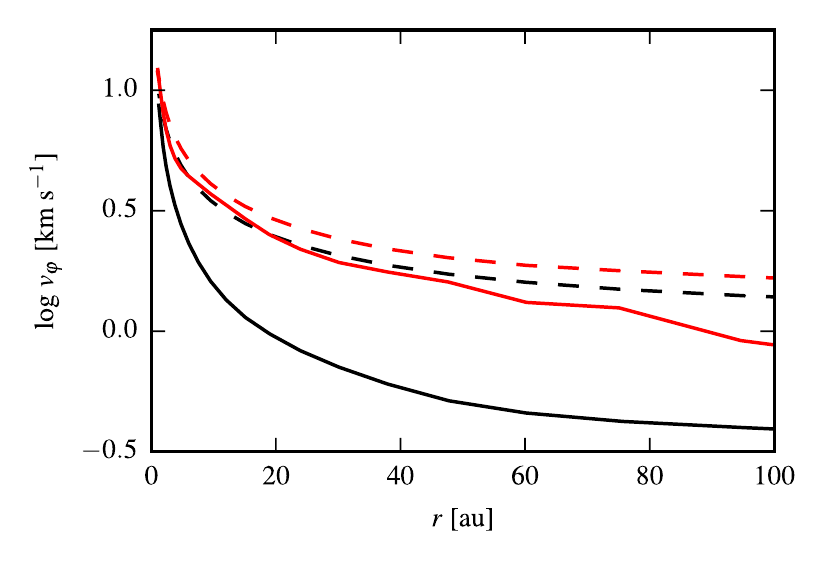}\vspace{-0.5cm}
\caption{\label{fig:vprof} Average tangential velocity ($v_\varphi$) as a function of radius in the pseudo-disc (defined here as the region $10\degree$ above and below the $x$--$y$ plane centered on the sink particle) at $t=1.02\tff$ for the $\vartheta = 0\degree$ $\mu = 5$ calculation (black solid line) and the $\mu = 10$ calculation (red solid line); the corresponding dashed lines show the Keplerian velocity at the same radius. The weaker magnetic field in the $\mu = 10$ calculation and the corresponding reduction in angular momentum transport from magnetic braking produce a faster rotation profile, promoting the formation of both a larger pseudo-disc and a faster bipolar outflow. Both rotation profiles are sub--Keplerian throughout, although at very small radii they approach the Keplerian velocity.}
\end{figure}

The subsequent evolution of the $\mu = 10$ calculations follows a similar overall pattern to $\mu = 5$ calculations. Those with $\vartheta \leq 45\degree$ produce collimated outflows, which are faster and more closely aligned to the rotation axis for lower $\vartheta$. As we discussed earlier, however, the reduction in the magnetic field strength and the \diff{consequently} reduced magnetic pressure support causes the core to collapse to higher densities more quickly.
\diff{A more rapid collapse results in a correspondingly earlier formation of the outflow and jet; whilst the reduced magnetic braking }
produces different pseudo-discs and outflow velocities compare to $\mu = 5$.

In the $\mu = 5$ calculations we obtained a pseudo-disc with a radii of $\approx 50~\au$ for $\vartheta < 45\degree$ and $< 15~\au$ for $\vartheta \geq 45\degree$. \Cref{fig:mu10all} (central columns) shows how for $\mu = 10$ the same initial values of $\vartheta$ produce much larger pseudo--discs: we obtain radii of $\approx 100~\au$ for $\vartheta \leq 45\degree$ and $25~\au$ for $90\degree$. We see the same correlation between pseudo-disc radius and $\vartheta$, whereby the disc radii reduce as the initial inclination of the field axis to the rotation axis is increased, but for $\mu = 10$ these are in general larger than the corresponding $\mu = 5$ calculation. The apparently larger discs seen in the left column of \cref{fig:mu5all}, particularly for $\vartheta = 0\degree$ compared to \cref{fig:mu10all} is due to a projection effect caused by looking along the plane of the disc, which due to preferential collapse of the the fluid parallel to the field lines has a higher density than the rest of the collapsed core. Our calculation does not include a physical viscosity treatment, therefore the main process for transporting angular momentum --- and hence allowing material to move towards the centre of the pseudo-disc --- is magnetic braking. 
\diff{The calculations do include artificial viscosity, but this is limited using the }\citet{1997JCoPh.136...41M}\diff{ switch and should produce negligible angular momentum transport compared to that produced by the magnetic field.}
The degree of magnetic braking increases with the magnetic field strength, $|B^i|$, and $\mu = 10$ implies an initial field which is half the magnitude of the $\mu = 5$ calculations, causing the formation of the large discs. As we noted for $\mu = 5$, the innermost region of the pseudo-disc (which drives the toroidal field and hence the collimated jet component of the outflow) realigns perpendicular to the rotation axis, although it is correspondingly larger for these calculations.

The most important difference between the two field strengths is that the outflows produced with $\mu = 10$ are $60~\%$ faster. This effect is most pronounced for the $\vartheta = 0 \degree$ calculation. Like the corresponding calculation in \cref{sec:discs:5}, this system initially had $|v_z| = 5~\kvelo$ with the characteristic collimated jet. However, the inner jet region detaches and a region of faster flowing material with $|v_z| = 8~\kvelo$ forms with the plasma $\beta$ above and below the pseudo--disc reducing from $10^{-2}$ to $10^{-3}$.

\Cref{fig:TorPol_1} shows the ratio of the poloidal to toroidal field for this calculation.
Whilst the $\mu = 5$ calculation had a weak poloidal component which was then tightly would by the toroidal field from the disc; as the cross--section in \cref{fig:TorPol_1} shows this calculation has a much stronger poloidal component, covering a larger region, above and below the disc. The disc is still dominated by the torodial \enquote{winding} field, but the extra pressure contributed by the poloidal region acts to detach the collimated component of the jet and increase the overall speed. This effect, which manifests as an extra (magnetic) pressure contribution, causes the reduced plasma $\beta$ seen. A similar, although weaker, effect causes the partial disconnections seen when $20\degree \leq \vartheta \leq 45\degree$.

We defined $B_p$ in \cref{sec:discs:5} as essentially the magnetic pressure. The cause of this extra field is then simply the more rapidly rotating pseudo--disc, as shown in the radially average $v_\phi$ plot in \cref{fig:vprof}, due to the initially reduced magnetic braking. In essence a larger disc and a faster outflow velocity are linked because the maximum jet velocity is itself linked to the maximum pseudo--disc rotational velocity, and this effect is continued across a range of misalignments. Additionally, because the initial field is weaker in these calculations, the $\vartheta = 45\degree$ calculation --- which for $\mu = 5$ produced a notably weaker jet than for $\vartheta = 20\degree$ --- has a more comparable velocity and structure, and a similarly larger inner disc region.

\subsection{$\mu = 20$}
\label{sec:discs:20}

\begin{figure*}
\centering{}
\includegraphics{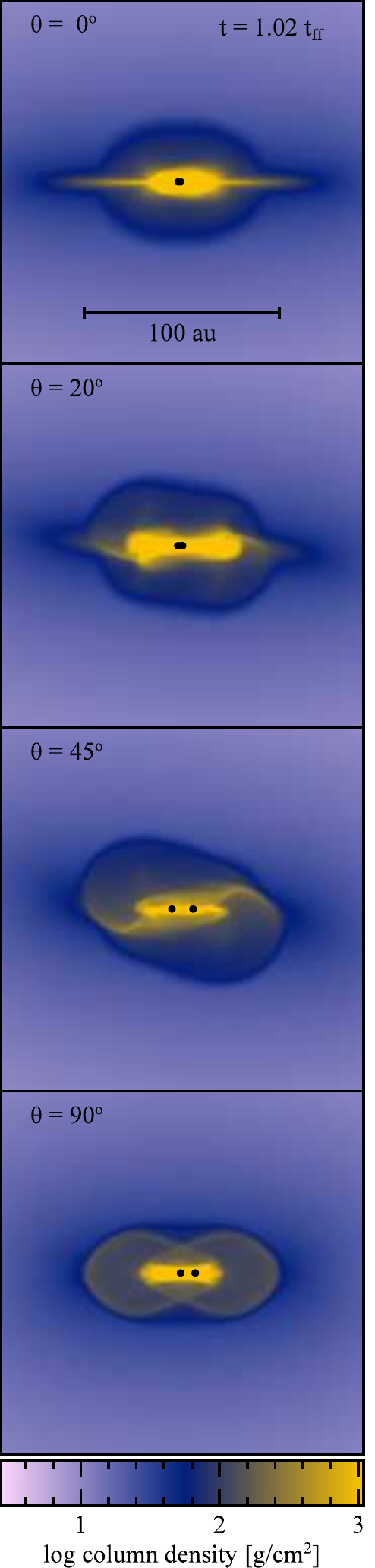}\includegraphics{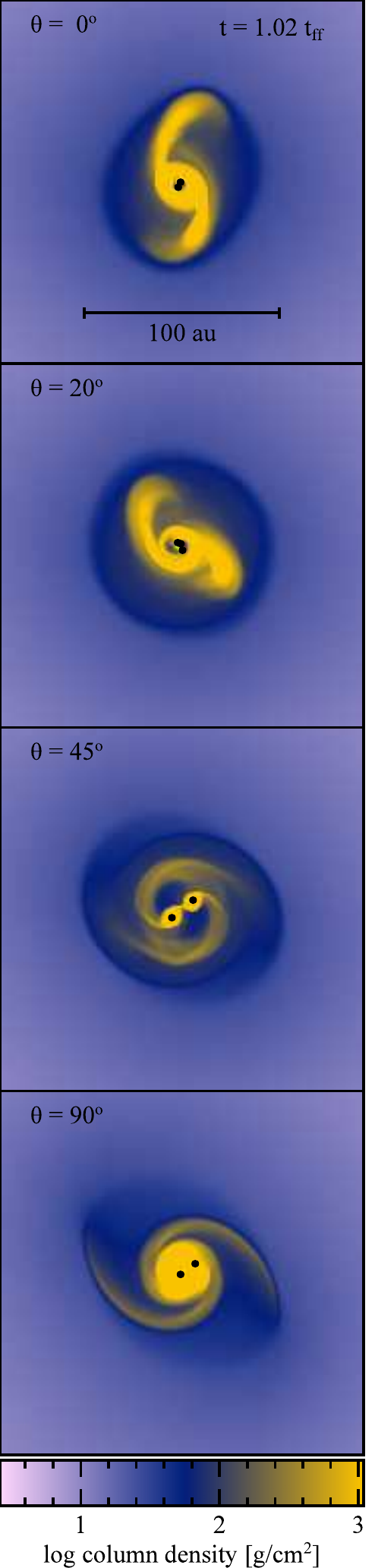}\includegraphics{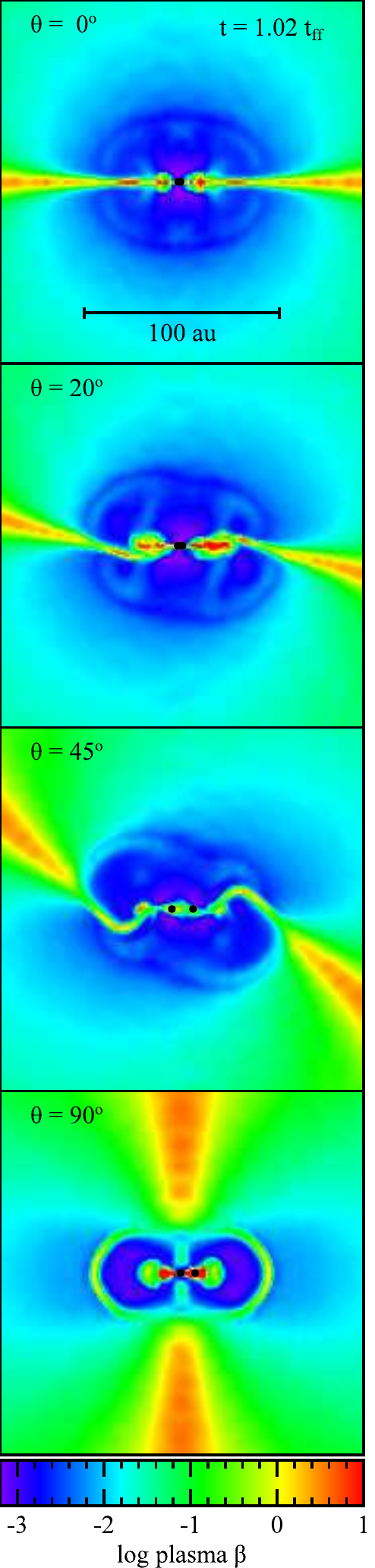}

\caption{\label{fig:mu20all} Column density projections in the direction of the $y-$axis (left
 column) and $z-$axis (centre column), and cross--sections of the plasma $\beta$ in the $z-x$ plane
  for the four $\mu = 20$ calculations at $t = 1.02\tff$. 
  The initial inclination between the 
  rotation and magnetic field axes decreases down the page, with the first row being $\vartheta = 0\degree$, 
  then $20\degree$, $45\degree$, and the bottom row having $\vartheta = 90\degree$. 
  \diff{Sink particles are represented by a black dot, slightly larger than the accretion radius.}
  The markedly different evolution for the weakest magnetic field calculations, compared to those shown in \cref{fig:mu5all,fig:mu10all}, including the formation of binary or multiple systems can be seen. We note that increasing $\vartheta$ increases the binary separation, and that at lower values both protostars are co-located within one disc, whilst at higher values a pair of discs are embedded in a circumbinary disc.
  }
\end{figure*}

\begin{figure}
\centering{}
\includegraphics{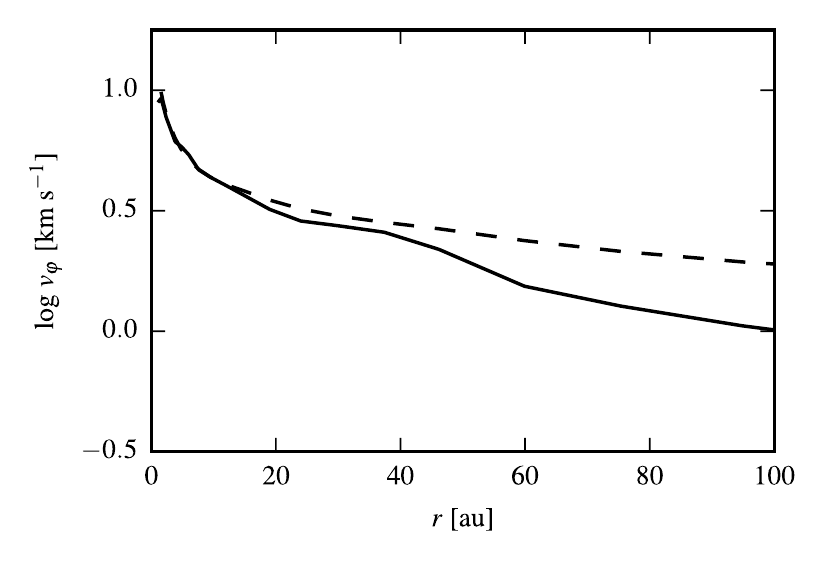}\vspace{-0.5cm}
\caption{\label{fig:vprof20} Average tangential velocity ($v_\varphi$) as a function of radius in the pseudo-disc, calculated in the same way as \cref{fig:vprof} but centered on the barycentre of the system rather than one of the protostars, at $t=1.02\tff$ for the $\vartheta = 0\degree$ $\mu = 20$ calculation (black solid line) with the corresponding black dashed line showing the Keplerian velocity profile. Compared to \cref{fig:vprof} the significantly reduced magnetic braking in this calculation allows the disc to become Keplerian within $r < 12~\au$ of the system barycentre.}
\end{figure}

\begin{figure*}
\centering{}
\includegraphics[width=\textwidth]{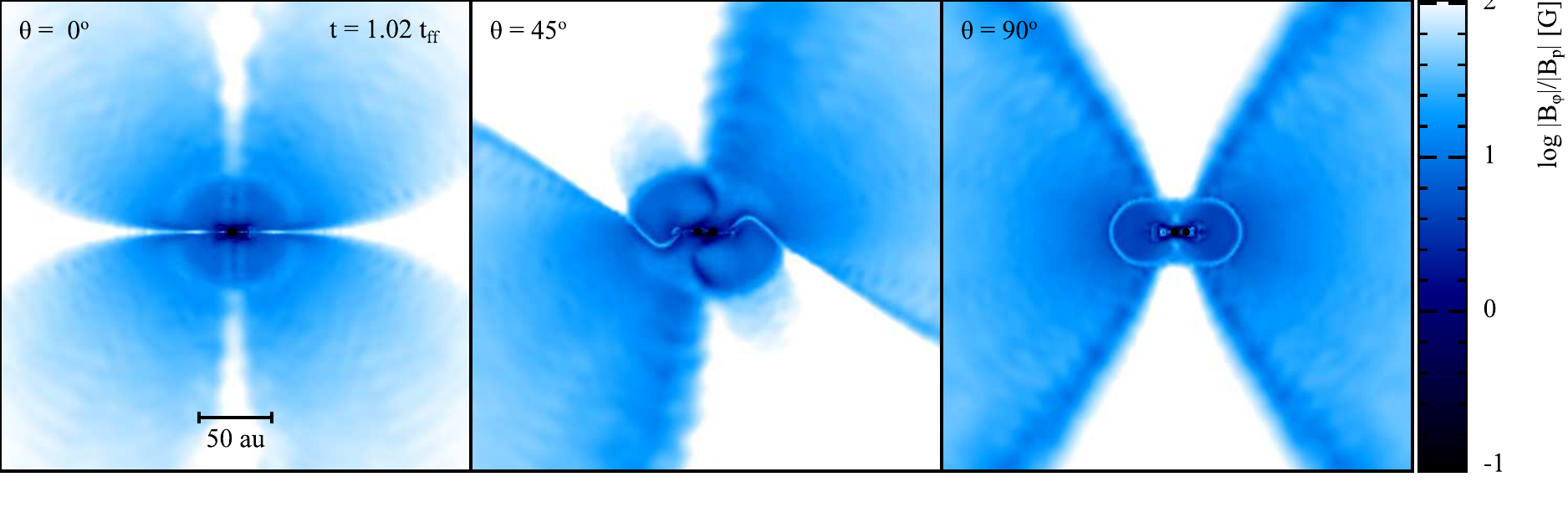}\vspace{-1.0cm}
\caption{\label{fig:TorPol_2} Cross--sections of the ratio of the toroidal field component to the poloidal magnetic field component (cf. \cref{eqn:radialB,eqn:polB,eqn:torB}) for $\vartheta = 0\degree$, $45\degree$, $90\degree$, $\mu = 20$ calculations at $t = 1.02\tff$, similar to \cref{fig:TorPol_0}. Values $<1$ indicate the poloidal component and conversely $>1$ indicates the toroidal component dominates.}
\end{figure*}
\begin{figure}
 \centering{}
 \includegraphics[width=0.45\textwidth]{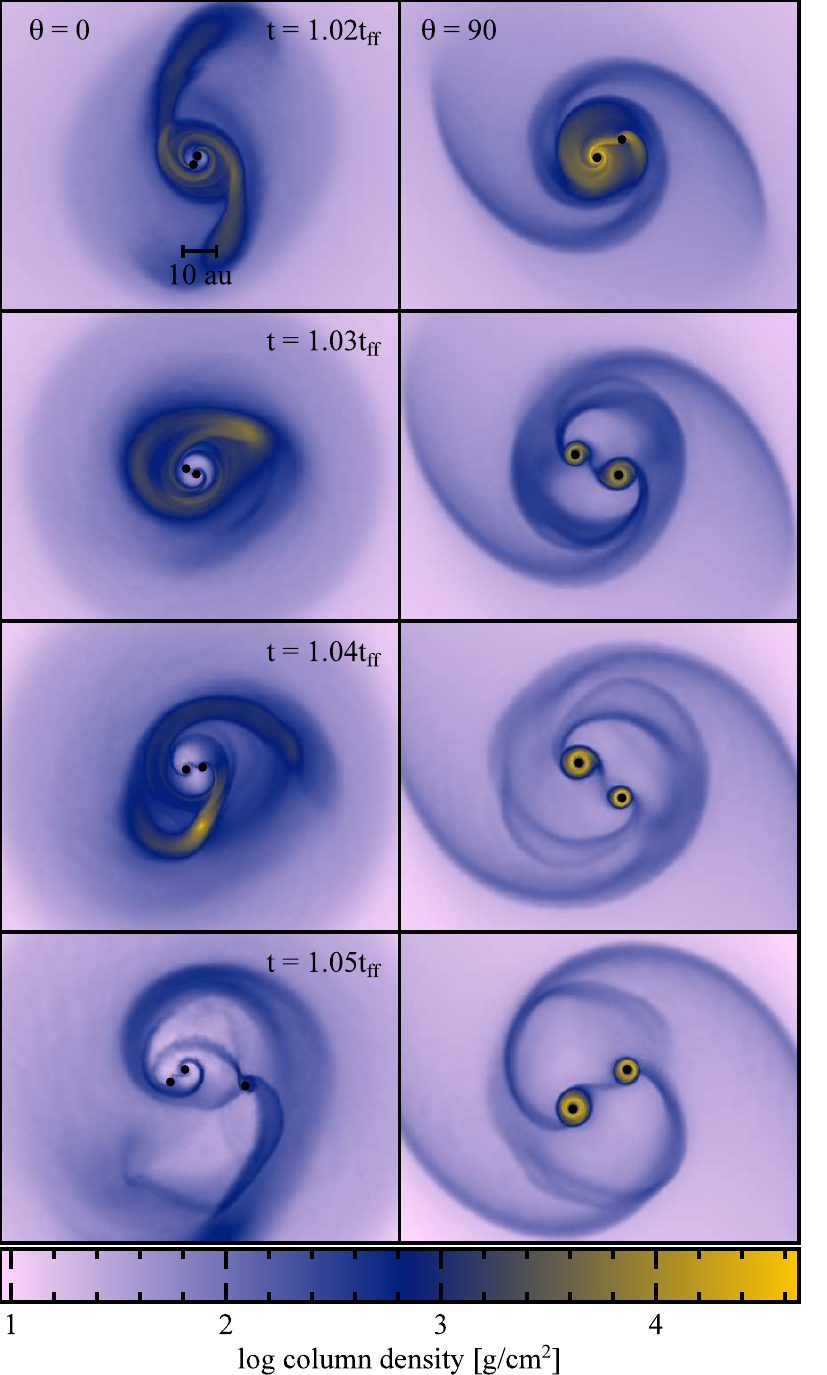}
 
 \caption{Column density projections for $\mu = 20$, $\vartheta = [0\degree, 90\degree]$ between $t = 1.02$ and $1.05\tff$. 
 \diff{Sink particles are represented by a black dot, slightly larger than the accretion radius.} 
 Both models have formed two sink by $1.02\tff$ but in a very different configuration: surrounded by one large disc for $0\degree$ compared to two smaller, separate, discs when $90\degree$. The $0\degree$ model eventually forms a third sink $\approx 10$ au away from the original pair.
 \label{fig:sinkpos}}
\end{figure}

\Cref{fig:mu20all} shows the results from the $\mu = 20$ calculations.
These do not form bipolar outflows instead fragmenting into binary (or multiple) systems, as seen in the central column of \cref{fig:mu20all}. The precursor to this is the formation of a small (ca. $10~\au$ radius) dense rotating disc: without the magnetic braking (and angular momentum loss from the outflow).
These smaller discs, which are closer in nature to true accretion discs (cf. \cref{fig:vprof20}) than the pseudo-discs discussed earlier, are gravitationally unstable. This can be seen from the distinct spiral arm-like structures.

The $\vartheta = 0\degree$ and $20\degree$ calculations form tight binaries with separations of $< 5~\au$. We add a cautionary note that in a calculation including additional physical processes, for example a radiative transfer scheme like that in \citet{2004MNRAS.353.1078W,2006MNRAS.367...32W}, these may not be formed due to the excess thermal support acting to stabilise the disc; alternatively they may merge together. In contrast, the separation between the two protostars for $\vartheta = 90\degree$ is $\approx 13~\au$ at $1.03\tff$. In all cases, the binary pairs are surrounded by a massive circumbinary disc, with a radius of $50~\au$, while the wider binaries have two distinct dense circumstellar discs (with radii of $< 5~\au$) embedded within this. 

All four values of $\vartheta$ produce an initial protostar by $t= 1.01\tff$, which is comparable to the time frame for a $\mu = 10$ calculation. However, these calculations then go on to produce a second protostar by $1.02\tff$, forming the binary pairs seen in \cref{fig:mu20all}. In addition, the $\vartheta = 0\degree$ model eventually produces a further sink particle from the circumbinary material, shown in \cref{fig:sinkpos}. The ultimate fate of this third potential protostar is unclear --- whilst a stable triple star system is possible, by  three-body interaction it could be ejected from the system or the two closer sinks could merge leaving a system similar to the $90\degree$ model albeit without the two discs). We note that this is a similar formation process to that proposed for the ternary system observed by \citet{2016Natur.538..483T}, albeit with a much smaller separation between the binary pair and the younger third protostar. However, an important difference between all the $\mu = 20$ models and the stronger fields is the absence of an outflow --- without which angular momentum can not be rapidly removed from the disc. This reduction in angular momentum transport means that fragmentation into a binary (which can store more angular momentum) is the result.

Although we noted earlier that these four calculations do not produce a bipolar outflow, there are still regions with substantial magnetic fields. Consequently a significant magnetic pressure is still realised and this produces a region around the circumstellar (or circumbinary) disc with $\beta_\rmn{plasma} < 1$. This causes some material to be moved away from the plane of the disc, although not in the rapid and directed manner of a jet. We see, for example, in \cref{fig:TorPol_2} that although a small poloidal region exists near the protostar, there is no collimating toroidal component present. This is more similar to the field structure seen for $\vartheta = 90\degree$ in \cref{fig:TorPol_0,fig:TorPol_1}, which similarly have no bipolar outflow and no significant toroidal component near the protostar.

In principle, this may indicate that binary stars may be formed by fragmentation even in non--turbulent clouds, or clouds which have not been perturbed \citep[\eg by an external impulse as in][]{1989MNRAS.239..361P} and are no longer axisymmetric; provided, of course, that the field is sufficiently weak so as to not drive a strong outflow. In practice, this may not be as strong a constraint as it seems --- in our quasi--ideal MHD the only method of dissipating magnetic energy is by artificial resistivity, which is intentionally limited to the minimum required for stability. In reality, non--ideal effects (resistivity, ambipolar diffusion, and the Hall effect) will act to reduce or transform the field and may therefore allow an effect similar to that seen here to occur for stronger initial fields. \citet{2016MNRAS.457.1037W} propose a consistent SPMHD method that extends the ideal MHD presented here to include all three non--ideal effects, and find that the Hall effect and the relative orientation of the rotation and magnetic field axes can have a significant effect on the consequent evolution of the core. Alternatively, as suggested in \citet{2005MNRAS.362..382M} it may be sufficient to just increase the initial angular momentum of the cloud to promote fragmentation at higher field strengths (or some combination of both mechanisms).

Conversely, there should exist a range of sufficiently strong initial fields so that no matter what other physical processes are at play, binary or multiple star systems are difficult to form, and hence solitary stars may be produced. Stronger fields naturally drive stronger jets and outflows or increased magnetic braking which transport angular momentum obviating the need for companion stars to store angular momentum.

\subsection{Accretion onto protostars}
\label{sec:accretion}

\begin{figure*}
\centering{}

\includegraphics{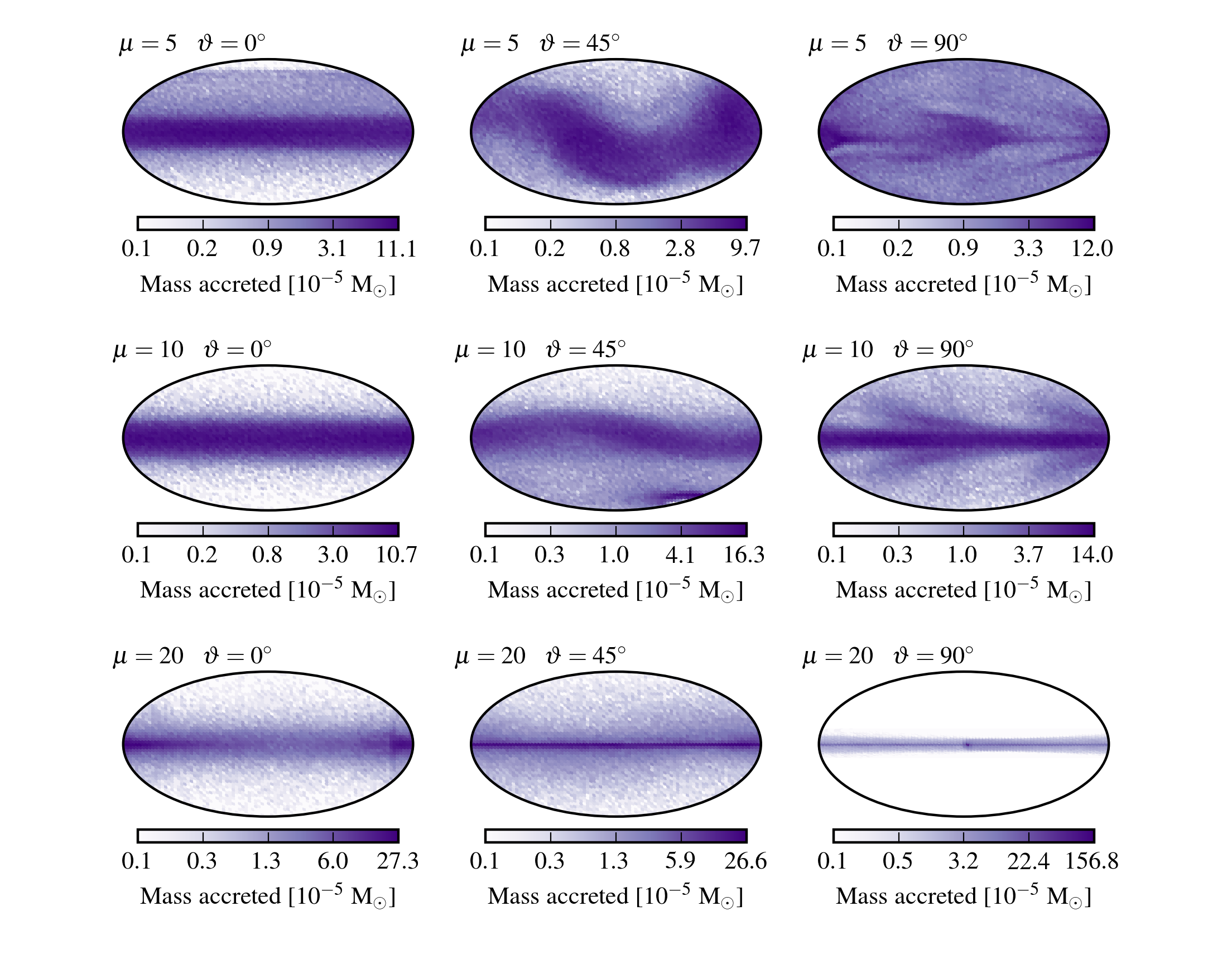}

\caption{
    Mollweide projections of the mass accreted onto a sink particle as functions of latitude and longitude during the first $500~\yr$ after formation. The top, middle and bottom rows are $\mu = 5$, $10$, and $20$ respectively, and the first, second and third columns are $\vartheta = 0\degree$, $45\degree$, and $90\degree$ (for $\mu = 20$ only the first sink particle is shown, a similar plot would be produced for the second particle). Accreted particles are placed in hexagonal bins and then projected into two dimensions using the Mollweide projection, the logarithm of the accreted mass is then used to set the colours.  All models preferentially accreted in the plane of their pseudo-disc, \eg the sinusoidal structure seen in two of the $45\degree$ models corresponds to a warped pseudo-disc.
    \label{fig:mollweide}
}

\end{figure*}

As well as producing outflows and discs with varying morphologies, each model also results in a different profile of accretion onto the sink particle (which represents a forming protostar). As before, a general trend is observed whereby the $\mu = 20$ models and $\vartheta = 90\degree$ models differ substantially from stronger fields and shallower angles. In \cref{fig:mollweide} we plot maps of the material accreted during the 500 yr after a sink particle is inserted. 

\begin{figure}
\centering{}

\includegraphics{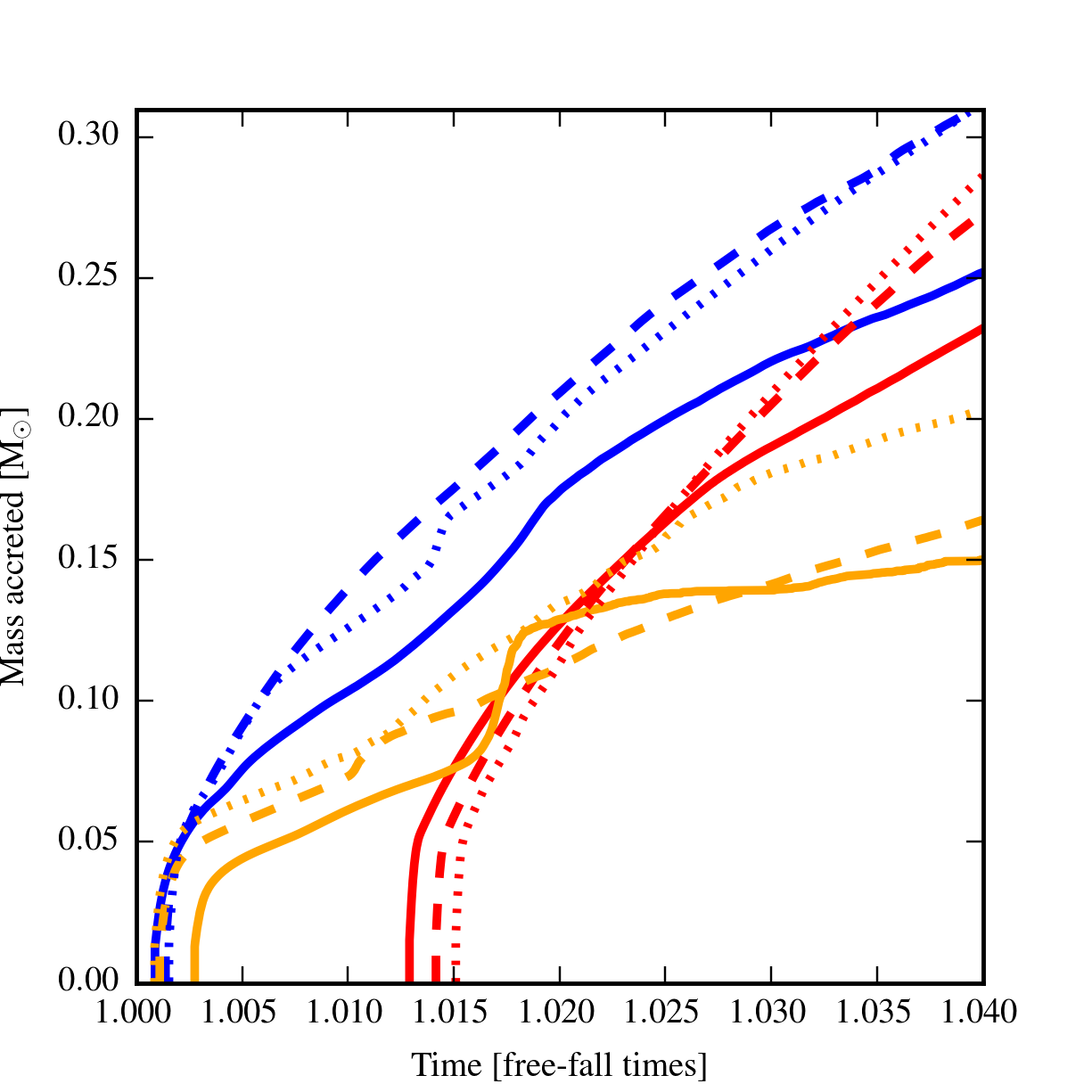}

\caption{
    Mass accreted by each sink particle plotted against time. The three magnetic field strengths are represented by red, blue and orange respectively for $\mu = [5,~10,~20]$ and solid, dashed and dotted lines represented $\vartheta = [0\degree,~45\degree,~90\degree]$. The later formation of the sink particle for $\mu = 5$ (see \cref{fig:sinklaunch}) at $t \approx 1.015 \tff$ is clearly shown. Additionally, we observe that whilst the $\mu = 10$ models form earlier, 
    the total mass accreted for each field geometry is similar for $\mu = 5$ and $10$ at $t=1.04\tff$.
    Whilst the $\mu = 20$ models appear to be accreting less, all these models produce a second sink by $t = 1.02\tff$ and which rapidly accretes an approximately similar mass; in addition a sink-sink interaction for $\vartheta = 0\degree$ causes the sharp knee in that model.
    \label{fig:acctime}
}

\end{figure}

\Cref{fig:acctime} shows that the amount of material accreted onto the sink
is broadly similar (at $t = 1.04 \tff$) across all of the $\mu = 5$ and $10$ models, which is born out by the similar outflow morphology seen earlier in \cref{sec:discs}. 
All calculations 
principally accrete material in the plane of their pseudo--disc\diff{, as seen in }\cref{fig:mollweide}. When $\vartheta = 0\degree$ this clearly aligns with the equator\diff{. Similarly, for weak magnetic fields, where $\mu = 20$ (cf. }\cref{fig:mu20all}\diff{), the accretion disc forms perpendicular to the rotation axis for all field geometries. The bottom row of }\cref{fig:mollweide}\diff{ shows how this results in a strongly equatorial accretion profile.}  For $\vartheta = 45\degree$ \diff{in the stronger field calculations} a sinusoidal structure is seen in \cref{fig:mollweide}, which is the expected shape for a warped pseudo--disc like those seen in \cref{fig:mu5all,fig:mu10all}. 
The $\mu = 5$, $\vartheta = 90\degree$ has not produced a clear disc and has instead taken on a sigmoidal structure. This causes a hemispherically symmetrical and rotationally symmetrical accretion pattern corresponding to each \enquote{arm} of the sigmoid. An imprint of a similar structure is seen for $\mu = 10$ from the two spiral arms of the disc. However, unlike in \mylbp\, we find that the change to the small scale structure results in a roughly equivalent accretion rate for both $\vartheta = 45\degree$ and $90\degree$. 

\subsection{Comparison to observations}
\label{sec:obs}

Observations of young stellar objects (YSOs) substantially more evolved than a first core \citep{1969MNRAS.145..271L} are plentiful compared to detections of objects which, although they have formed some sort of core, are much less evolved. 
Detections of an earlier phase --- a dense starless core which is approaching the point of forming a star --- are better represented, \eg those found by \citet{2005ApJ...619..379C}. Therefore, constraints on the dynamics a first hydrostatic core and its environs would be useful.
The insertion of sink particles means we do not follow the collapse to exactly first hydrostatic core proposed by \citet{1969MNRAS.145..271L}. However, our sink particle has comparable dimensions so the overall structure of the pseudo--disc and any outflow should be similar.

In models where we obtain a collimated (or jet--like) outflow we find that the bulk velocity ranges between $\approx 2$ and $\approx 8~\kvelo$  depending on the exact configuration of parameters, with higher values of $\vartheta$ corresponding to generally reduced outflow velocities. This substantially increases the range of potential outflow velocities which may be first cores as opposed to \enquote{very low luminosity objects} \citep[VeLLOs, see][]{2005AAS...207.6353M,2005AN....326..878K}. One example is given in \citet{2011ApJ...743..201P}, who find that L1451-mm has a slow outflow with a velocity of about $3~\kvelo$. Similarly, \citet{2011ApJ...742....1D} observe an bipolar outflow of comparable velocity in Per--Bolo 58. These velocities are apparently slightly too slow if we only consider lower values of $\vartheta$, however, an outflow of this velocity would be approximately consistent with our results for $\vartheta = 45\degree$. 

At the other extreme, \citet{2010ApJ...715.1344C} find that L1448 IRS2E (another first core candidate) has a clearly collimated outflow with a velocity of around $25~\kvelo$. This is much faster than any of the models presented here, which have an artificial limit of $1~\au$ placed on the collapse. Outflow velocity and the maximum rotation speed (which is controlled by the larger of the size of the protostar or sink particle) are closely linked \citep{2003MNRAS.339.1223P} and consequently this indicates that whatever object is present in L1448 IRS2E, it has a substantially smaller radius than $1~\au$, and may therefore be more evolved than a first core.  

In effect we find that slow bipolar outflows are characteristic of first cores at all but the weakest field strengths. 
This implies that there is a strong constraint on first core candidates: any outflow, if present, should be between $\approx 2$ and $8~\kvelo$ and that faster outflows may be indicative of a more evolved object.

\section{Conclusions}
\label{sec:conclusions}

Using smoothed particle magnetohydrodynamics we have performed twelve simulations (using three mass--to--flux ratios, and four angles between the field and rotation axes) of the collapse of molecular cloud cores to form protostars. Whilst we observe that the initial collapse of the cloud core is broadly unaffected by the choice of parameters, the subsequent evolution can be quite different. All four $\mu = 20$ calculations produce binary systems of some form, with the degree of binary separation strongly dependent on the initial field geometry. At shallow angles, \ie low values of $\vartheta$ the two protostars form very close together, within $\lesssim 5 \text{~au} $; at steeper angles two separate discs form orbiting a common barycentre. This may indicate that binary or multiple star systems may be formed in molecular clouds if the initial magnetic field is very weak. We also find that at this field strength significant first core outflows are absent: instead the system conserves angular momentum by forming a binary.

In contrast, no simulation with $\mu \leq 10$ formed a binary for any field geometry. However, the structure and nature of the resultant pseudo--disc and outflow varies between both values of $\mu$ and $\vartheta$. In addition, the $\mu = 10$ models are observed to collapse somewhat more quickly than those with a stronger field. We find that for all values of $\vartheta$, the pseudo--disc formed for $\mu = 10$ is significantly larger than for $\mu = 5$ due to reduced magnetic braking. In particular, when $\vartheta = 90\degree$ the stronger field fails to produce a disc at all, compared to the weaker field where a comparatively large disc. Additionally, the outflows produced may have different speeds, ranging from $2$ to $8~\kvelo$ depending on $\mu$, $\vartheta$, and time.

Additionally, changing the initial field geometry from a configuration where the rotation and field axis are aligned to one where they are misaligned can promote the formation of warped pseudo--discs. Here the outer regions of the discs orient perpendicular to the magnetic field axis (due to reduced magnetic pressure support parallel to the field lines) but the inner region begins to re--orient perpendicular to the rotation axis as angular momentum is concentrated near the sink particle.

These results may have consequences for how binary or single star systems are formed. Clearly, a weaker field, with the attendant lack of both magnetic braking and outflow, is liable to fragment into a binary. Similarly, at the strongest field strengths (with aligned fields), the effect of the magnetic braking plus a large collimated outflow may be sufficient to prevent this fragmentation from occurring. However, a large intermediate regime exists which may allow the formation of binary or single star systems depending on the exact parameters of the initial cloud core; although adding additional physics, for example radiative transfer, may affect this. Further work, following the collapse to a protostellar core itself would be necessary to determine the exact boundary between the two regimes.

\section*{Acknowledgments}
The authors thank D. J. Price for his comments on the draft manuscript.

BTL acknowledges support from an STFC Studentship and Long Term Attachment grant. 
This work was also supported by the European Research Council under the European Community's Seventh Framework Programme (FP7/2007-2013 Grant Agreement No.
339248). MRB's visit to Monash was funded by an International Collaboration Award from the Australian Research Council (ARC) under the Discovery Project scheme grant DP130102078. 

This work used the DiRAC Complexity system, operated by the University of Leicester IT Services, which forms part of the STFC DiRAC HPC Facility (www.dirac.ac.uk). This equipment is funded by BIS National E-Infrastructure capital grant ST/K000373/1 and STFC DiRAC Operations grant ST/K0003259/1. DiRAC is part of the National E-Infrastructure.

Calculations were also performed on the University of Exeter Supercomputer, a DiRAC Facility jointly funded by STFC, the Large Facilities Capital Fund of BIS and the University of
Exeter.

Rendered plots were produced using the \textsc{splash} \citep{2007PASA...24..159P} visualisation programme. 

\bibliographystyle{mnras}
\bibliography{Outflows_weak_fields}

\bsp

\label{lastpage}
\end{document}